\documentclass[reprint,amssymb, amsmath, aps, superscriptaddress, showpacs, footinbib, prl]{revtex4-2}
\usepackage{graphicx,epstopdf} %Include figure files,superscriptaddress
\usepackage{amsmath}
\usepackage{xcolor, soul}

\usepackage[colorlinks,bookmarks=false,citecolor=black,linkcolor=black,urlcolor=black]{hyperref}
\usepackage{amssymb}
\usepackage{longtable}
\usepackage{soul}
\newcommand{\be}{\begin{equation}}
\newcommand{\ee}{\end{equation}}
\newcommand{\bea}{\begin{eqnarray}}
\newcommand{\eea}{\end{eqnarray}}
\newcommand{\bse}{\begin{subequations}}
	\newcommand{\ese}{\end{subequations}}

\begin{document}

\begin{titlepage}
\title{ Higher order exchange driven noncoplanar magnetic state and large anomalous Hall effects  in electron doped kagome magnet Mn$_3$Sn  } 

\author{Charanpreet Singh}
\affiliation{School of Physical Sciences, National Institute of Science Education and Research, An OCC of Homi Bhabha National Institute, Jatni-752050, India}

\author{Sk Jamaluddin}
\affiliation{School of Physical Sciences, National Institute of Science Education and Research, An OCC of Homi Bhabha National Institute, Jatni-752050, India}

\author{Ashis K. Nandy}
\affiliation{School of Physical Sciences, National Institute of Science Education and Research, An OCC of Homi Bhabha National Institute, Jatni-752050, India}

\author{Masashi Tokunaga}
\affiliation{The Institute for Solid-State Physics, University of Tokyo, 5-1-5 Kashiwanoha, Kashiwa, Chiba, 277-8581, Japan}

\author{Maxim Avdeev}
\affiliation{Australian Nuclear Science and Technology Organisation, New Illawarra Road, Lucas Heights, New South Wales 2234, Australia}

\author{Ajaya K. Nayak}
\email{ajaya@niser.ac.in}
\affiliation{School of Physical Sciences, National Institute of Science Education and Research, An OCC of Homi Bhabha National Institute, Jatni-752050, India}

\begin{abstract}
Owing to the geometrical frustration, Mn$ _3 $Sn exhibits a 120$^{\circ}$ in-plane triangular antiferromagnetic (AFM) order with a large anomalous Hall effect (AHE). Here, we present a combined theoretical and experimental study to demonstrate that the in-plane AFM structure in Mn$ _3 $Sn can be significantly modified to a tunable  noncoplanar magnetic state by suitable electron doping. With the help of Density Functional Theory calculations we show that the presence of higher-order exchange interactions in the system leads to the stabilization of the noncoplanar magnetic ground state, which is further established by neutron diffraction study in the Fe-doped Mn$ _3 $Sn sample. Interestingly, we find a large scalar spin chirality  (SSC) induced topological AHE that  can be significantly tuned with the degree of non-coplanarity. We carry out 60~T magnetic and Hall resistivity measurements to demonstrate the contribution of SSC to the observed AHE. We also illustrate a simultaneous manipulation of dual order in the system, where the AHE arising from the in-plane 120$^{\circ}$ triangular AFM order can switch its sign without affecting the AHE generated by the SSC.   The present study opens up a new direction to explore novel quantum phenomena associated with the coexistence of multiple orders.

\end{abstract}
\maketitle
	
\end{titlepage}	

%\section*{\textbf{Introduction}}

In general, the magnetically ordered states are the manifestation of exchange interactions within the Heisenberg model, which accounts for a variety of magnetic orders ranging from  simple collinear  to complex non-collinear magnetic states. However, the energy landscape of magnetic interactions are not limited to the usual pairwise Heisenberg exchange ($J$), instead higher-order exchange terms corresponding to multiple order of hopping between different sites can be taken into account in suitable magnetic systems. The most notable interaction involving multi-spins in a magnetic Hamiltonian is the  4-spin exchange in the form of 4-spin-two-site ($B$), 4-spin-three-site ($Y$), and 4-spin-four-site ($K$) interactions (Eq. \ref{Hemil}) \cite{HOE_Adler_1979}. A signature feature of the 4-spin exchange interactions is the stabilization of $multi$-$Q$ states. The noncoplanar $3Q$ magnetic ordering detected in  Mn/Cu(111) system, and the unique $2Q$  ground state observed in Fe/Rh(111)  can be well explained by considering the 4-spin exchange interaction \cite{3Q_exp_PRL,uudd_exp_3spin,systematicDerivation_highOrder}. In addition, certain  noncoplanar spin structures can be characterized by scalar spin chirality (SSC), defined as $\bf S_i . (S_j\times S_k)$,  which gives rise to an additional Hall response owing to the real space Berry phase. Such extraordinary Hall response has been previously observed in glassy systems \cite{AHE_spinglass_theory,AuFeHallSSC,AuFeHallSSC2,AHE_nagaosa_review} and a limited number of pyrochlore lattices \cite{AHE_SSC_toura_science,AHE_SSC_withoutdipole,AHE_spinglass_aufe,AHE_nagaosa_review}.

%Hence, it is important to explore the effect of higher order exchange interactions on the magnetic ground state of different lattice systems. 

%%%%%%%%%%%%%%%%%%%%%%%%%%%%%%%%%%%%%%%%%%%%%%%%%%%%%%%%%%%%%%%%%%%%%%%%%%%%%%%%%%%%%%%%%%%%%%%%%%%%%%%%%%%%
\begin{figure*}[tb]
	\begin{center}
		\includegraphics[angle=0,width=\textwidth,clip=true]{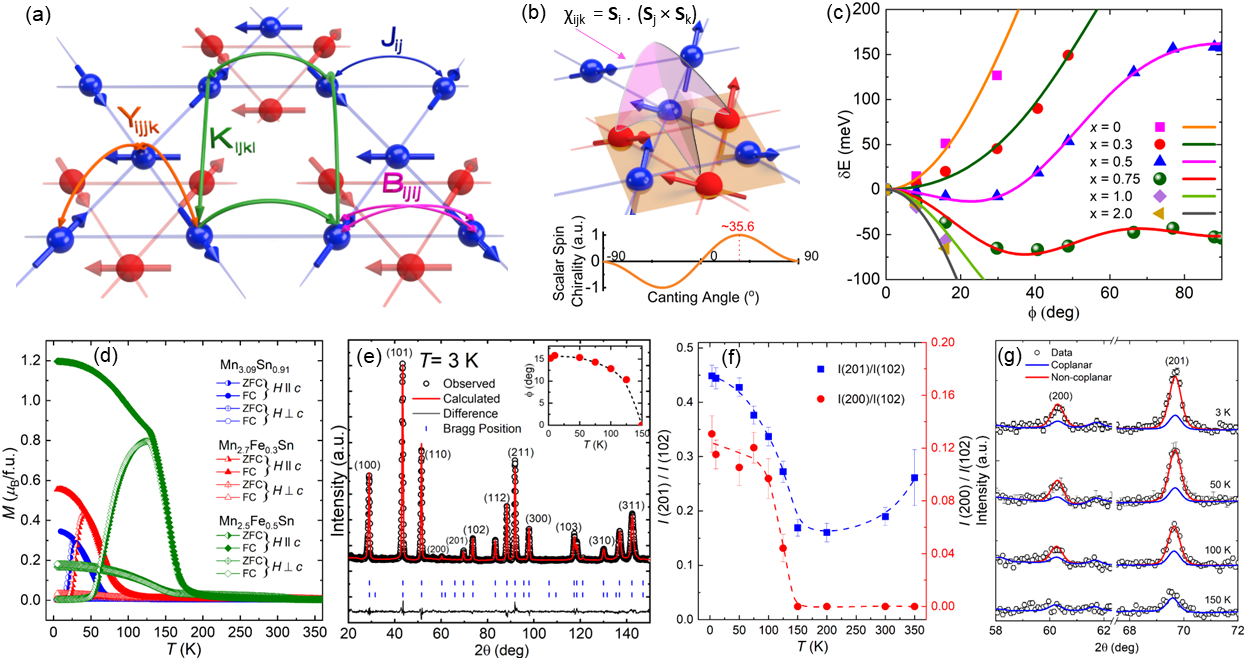}
		\caption{\label{Fig1} (a) Schematic diagram of the 120$^{\circ}$ in-plane triangular AFM structure of Mn$ _3 $Sn.  The red and blue atoms represent two layers of kagome lattice formed by Mn atoms. Hopping paths for $J_{ij}$, $B_{ij}$, $Y_{ijjk}$, and $K_{ijkl}$  are shown by blue, magenta, orange, and green curved arrows, respectively.  (b) Schematic  of an out-of-plane canted magnetic state with a finite scalar spin chirality (SSC), shown by the solid cone. The variation of SSC with canting angle is plotted at the bottom. (c) $ab$ $initio$ calculated change in energy (solid symbols) for Mn$_{3-x}$Fe$_{x}$Sn as a function of out-of-plane canting angle $\phi$.  The lines represent  fittings with  equation \ref{Hemil}. (d) Temperature-dependent magnetization  for  Mn$_{3-x}$Fe$_{x}$Sn samples with field parallel and perpendicular to the c-axis. (d) Rietveld refinement of the neutron diffraction (ND) pattern at 3~K for Mn$_{2.5}$Fe$_{0.5}$Sn. The inset shows extracted canting angle  at different temperatures. (e) Change in the relative integrated intensity of (200) and (201) peaks with respect to the (102) peak as a function of temperature. Dotted lines serve as a guide to the eye. (f) Expanded view of the (200) and (201) ND patterns at different temperatures, along with the Rietveld refinement for coplanar and noncoplanar models. }
	\end{center}
\end{figure*}

%%%%%%%%%%%%%%%%%%%%%%%%%%%%%%%%%%%%%%%%%%%%%%%%%%%%%%%%%%%%%%%%%%%%%%%%%%%%%%%%%%%%%%%%%%%%%%%%%%%%%%%%%%%%
 
The Kagome lattice system, which can be constructed from the triangular lattice by  removing one of the lattice points from the unit cell, exhibits a great prospect  for the realization of various types of spin-ordering. Among them, Mn$_3$Sn is one of the extensively studied materials  due to its multiple magnetic ordered states that include a 120$^{\circ}$ triangular antiferromagnetic (AFM) state. The cluster octupole order of the triangular AFM state  breaks the time-reversal symmetry,  resulting in the observation of  Weyl nodes  \cite{mn3sn_octupole_theory,mn3sn_kerr,mn3sn_octupole_xray,Weyl_mn3Sn_theory}, which leads to the observation of large  anomalous Hall effect in Mn$_{3}$Sn and Mn$_3$Ge \cite{AHE_mn3ge_Exp,AHE_mn3sn_Exp_Nakatsuji}. Interestingly, Mn$_3$Sn displays a temperature-dependent modification of the magnetic ground state with different levels of electron doping \cite{mn3sn_phaseDependHall,mn3sn_pressure,mn3sn_kondo,mn3sn_samplequalityHall}. However, the octupole order remains as the ground state at room temperature for all the electron-doped Mn$_3$Sn samples. Below the room temperature, the existence of a helical modulation of the triangular spin structure along the $c$-axis has been reported for the near stoichiometric Mn$_3$Sn samples \cite{mn3sn_phaseDependHall,mn3sn_excitation_park,mn3sn_pressure}.  For the higher electron-doped Mn$ _3 $Sn samples, the possible existence of a canted magnetic structure at very low temperature is discussed in a recent report \cite{mn3sn_THE_pradeep}. However, the exact nature and mechanism that governs the canted magnetic state and related transport phenomena remain unclear. %An combination of the typical pair wise Heisenberg exchange and single ion anisotropy can not explain the possible canted state. 
%Here we explore the low temperature ground state of Mn$_3$Sn in regard of higher order exchange interaction.
%%%%%%%%%%%%%%%%%%%%%%%%%%%%%%%%%%%%%%%%%%%%%%%%%%%%%%%%%%%%%%%%%%%%%%%%%%%%%%%%%%%%%%%%%%%%%%%%%%%%%%%%%%%%
%\textbf{Noncoplanar magnetic order with electron doping} : 

The 120$^{\circ}$ triangular AFM ground state of  Mn$_3$Sn can be derived from the  $J$ with in-plane magnetic anisotropy  \cite{mn3sn_excitation_park}. Possible higher-order exchange interactions along with the nearest neighbor Heisenberg exchange and  corresponding  hopping paths for  Mn$_3$Sn  are depicted  in Fig \ref{Fig1} (a). The  Hamiltonian of the system with the 4-spin interactions  can be written as,

\begin{multline} \label{Hemil}
	H = \sum_{ij} J_{ij}(S_i . S_j) + \sum_{ij} B_{ij}(S_i . S_j)^2 \\ + \sum_{ijk} Y_{ijk} [(S_i . S_j)(S_j . S_k)+(S_j . S_i)(S_i . S_k)+(S_i . S_k)(S_k . S_j)] \\ + \sum_{ijkl} K_{ijkl} [(S_i . S_j)(S_k . S_l)+(S_i . S_l)(S_j . S_k)-(S_i . S_k)(S_j . S_l)].
\end{multline}

%\begin{multline} \label{Hemil}
%	H = \sum_{ij} J_{ij}(s_i . s_j) + \sum_{ij} B_{ij}(s_i . s_j)^2 \\ + \sum_{ijk} Y_{ijk} [(s_i . s_j)(s_j . s_k)+(s_j . s_i)(s_i . s_k)+(s_i . s_k)(s_k . s_j)] \\ + \sum_{ijkl} K_{ijkl} [(s_i . s_j)(s_k . s_l)+(s_i . s_l)(s_j . s_k)-(s_i . s_k)(s_j . s_l)].
%\end{multline}
As discussed earlier, the magnetic ground state of Mn$_3$Sn can be effectively modified by a small electron doping. Motivated by these previous reports,  we carry out a detailed first principle electronic calculation for the electron-doped Mn$_3$Sn samples \cite{Supplemantary}.  Using virtual crystal approximation (VCA) technique as implemented in the Vienna Ab initio Simulation Package (VASP) code \cite{vasp1,vasp2,vasp3,vasp4},  we systematically vary the electron concentration by substituting  Fe in place of Mn in Mn$_{3-x}$Fe$_{x}$Sn.  Starting from the in-plane 120$^{\circ}$ structure, we calculate the total energy of the pristine and the electron-doped samples  by subjecting all the Mn moments with equal rotation from the kagome plane towards the c-axis, as represented in Fig. \ref{Fig1} (b). The change in energy as a function of the out-of-plane canting angle $\phi$  is plotted in Fig. \ref{Fig1} (c).  In the case of the undoped  Mn$_{3}$Sn, we find a minimum at $\phi = 0$, signifying an in-plane triangular AFM order as the ground state. The calculated energy curve can be fitted only with the pairwise Heisenberg exchange term \cite{Supplemantary}. A similar trend is found for $ x=0.3 $, albeit the stability of the triangular AFM state ($\phi$=0$^{\circ}$) reduces significantly in comparison to the FM order ($\phi$=90$^{\circ}$).   However, a satisfactory fit of the calculated energy for  $ x=0.3 $ can only be achieved with the inclusion of 4-spin exchange terms in addition to the 2-spin Heisenberg exchange.

%\textcolor{red}{We use the constrained magnetic moment calculation in Vienna Ab initio Simulation Package (VASP) to probe  the possible existence of noncoplanar magnetic ground state.}

%%%%%%%%%%%%%%%%%%%%%%%%%%%%%%%%%%%%%%%%%%%%%%%%%%%%%%%%%%%%%%%%%%%%%%%%%%%%%%%%%%%%%%%%%%%%%%%%%%%%%%%%%%%%
\begin{figure}[tb]
	\begin{center}
		\includegraphics[angle=0,width=8.5 cm,clip=true]{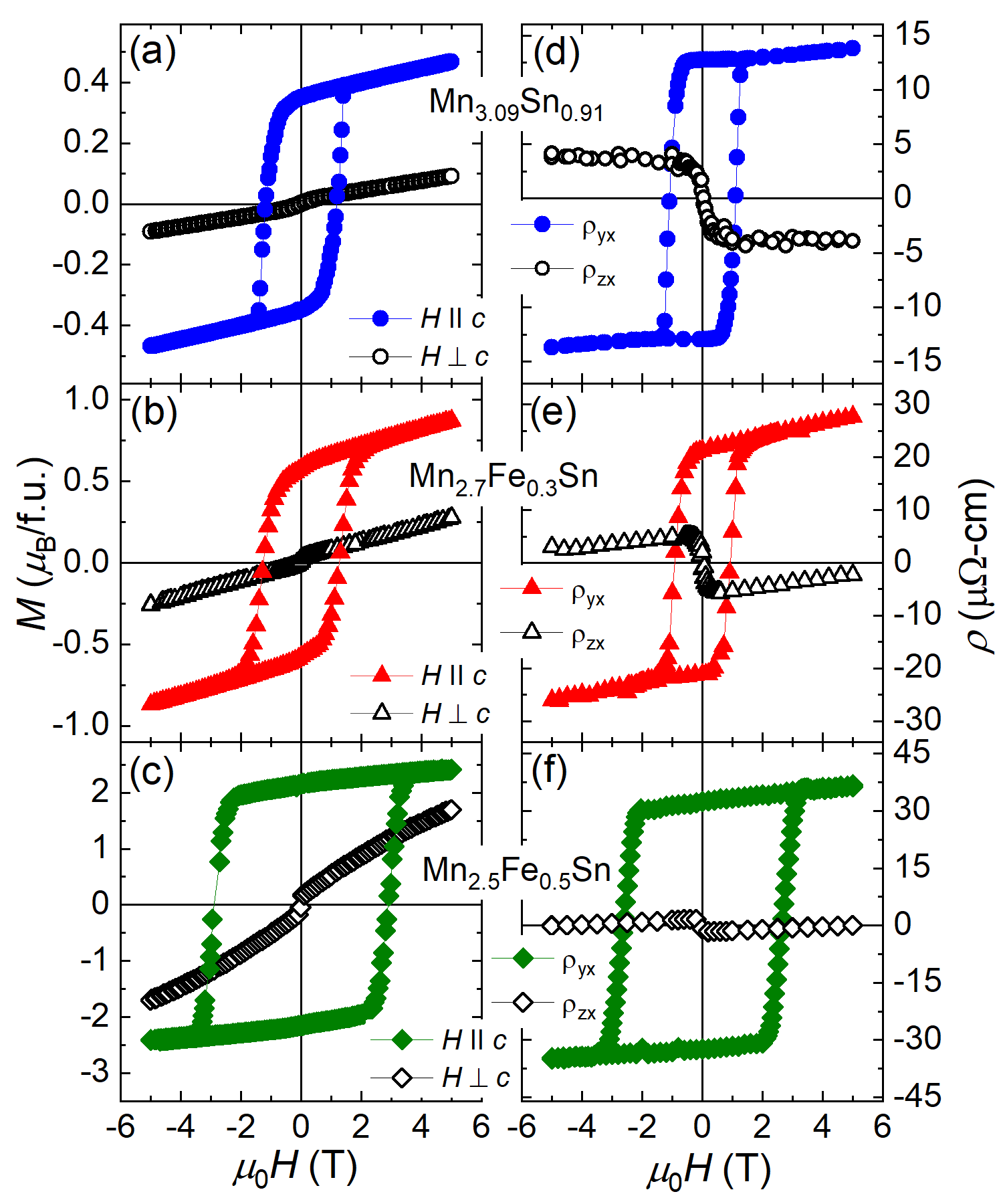}
		\caption{\label{Fig2}  Field dependent magnetization, $M (H)$, loops measured at 5~K with field parallel and perpendicular to the c-axis for (a) Mn$_{3.09}$Sn$_{0.91}$, (b) Mn$_{2.7}$Fe$_{0.3}$Sn, and (c) Mn$_{2.5}$Fe$_{0.5}$.  (d-f) Hall resistivity data,  $\rho_{yx}$ and $\rho_{zx}$ for the above three samples measured at $ T= $5~K. }
	\end{center}
\end{figure}
%%%%%%%%%%%%%%%%%%%%%%%%%%%%%%%%%%%%%%%%%%%%%%%%%%%%%%%%%%%%%%%%%%%%%%%%%%%%%%%%%%%%%%%%%%%%%%%%%%%%%%%%%%%%

 %%%%%%%%%%%%%%%%%%%%%%%%%%%%%%%%%%%%%%%%%%%%%%%%%%%%%%%%%%%%%%%%%%%%%%%%%%%%%%%%%%%%%%%%%%%%%%%%%%%%%%%%%%%%%%%%%%%%%%%%%%%%%%%%%%%%%%%%%%%%%
 \begin{figure}[tb]
 	\begin{center}
 		\includegraphics[angle=0,width=8.5 cm,clip=true]{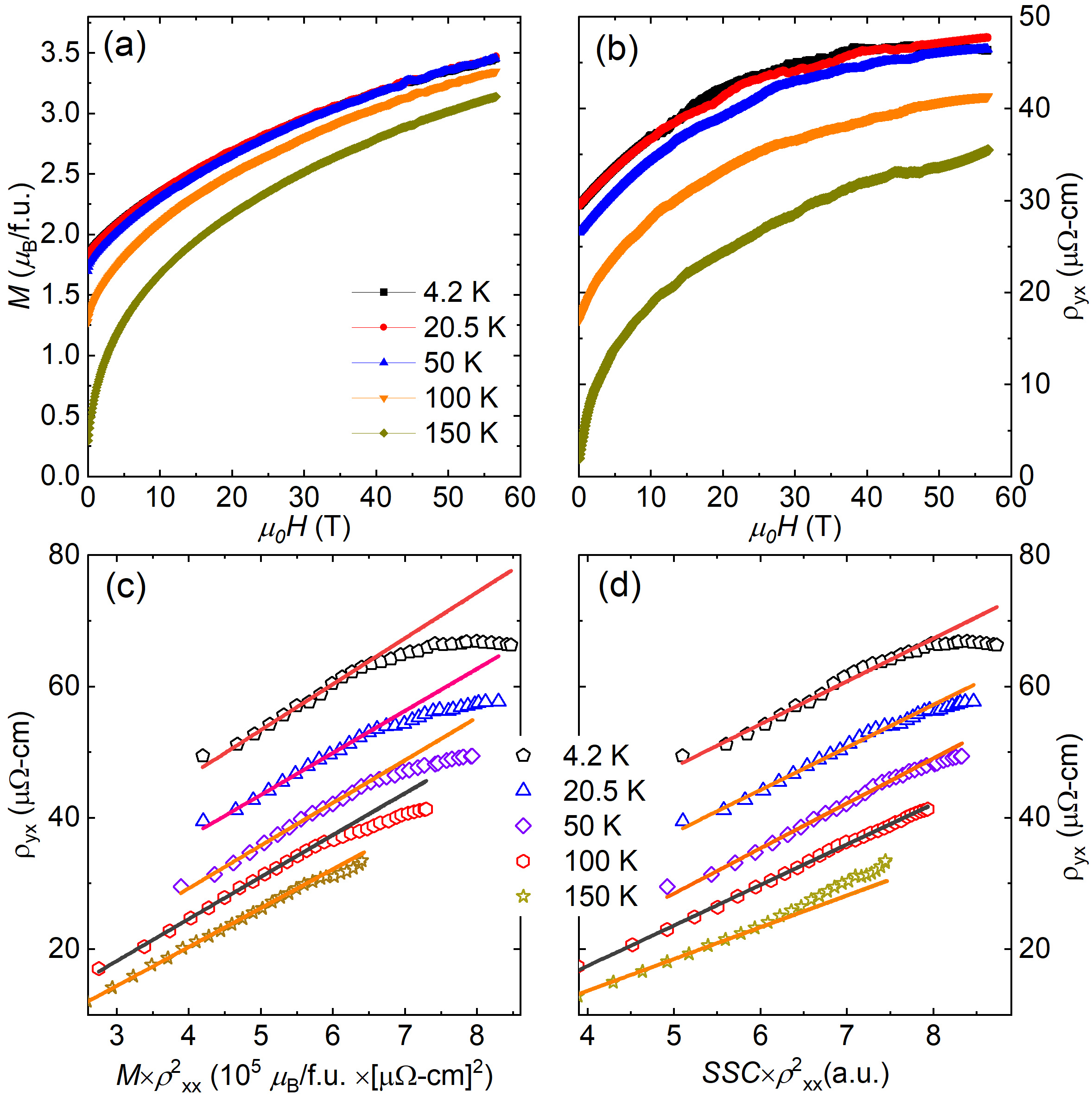}
 		\caption{\label{Fig3}  Field dependent (a) magnetization and (b) Hall resistivity data, $\rho_{yx}$  measured for fields up to 60~T. (c-d) Plot of $\rho_{yx}$ against magnetization $M \times \rho^2_{xx}$, and SSC$ \times \rho^2_{xx}$. The lines represent a linear fit. The data in (c-d) is shifted along the $y$-axis for better clarity. }
 	\end{center}
 \end{figure}
 %%%%%%%%%%%%%%%%%%%%%%%%%%%%%%%%%%%%%%%%%%%%%%%%%%%%%%%%%%%%%%%%%%%%%%%%%%%%%%%%%%%%%%%%%%%%%%%%%%%%%%%%%%%%
By further increasing the electron doping in  $ x=0.5 $, a drastic change in the nature of the energy curve is observed where a shallow minimum appears at a finite value of the out-of-plane canting angle with $\phi$=23$^{\circ}$ [Fig. \ref{Fig1} (c)]. The position of the minimum further increases to about $\phi$=37$^{\circ}$ for   $ x=0.75 $. For Fe doping beyond $1.0$, the FM state comes out to be the ground state. Note that the end compound Fe$_ 3 $Sn exhibits a FM ground state \cite{fe3sn}.  In contrast  to the $3Q$ magnetic state with zero net magnetic moment found in the case of  triangular lattice, here, the canted magnetic structure introduces a finite out-of-plane moment \cite{3Q_exp_PRL}.  The energy curves for the intermediate Fe doped samples are best fitted upon inclusion of 6-spin interaction in the form of $X_{ijk}(S_i . S_j)(S_j . S_k)(S_k . S_l)$ along with the 4-spin and 2-spin terms \cite{chiral-chiralExchange,chiralMultisite_Samir}. The extracted values of different order exchange constants from the fitting reveals that the canted magnetic ground state in the present case is stabilized by the higher order exchange contributions \cite{Supplemantary}. It is worth mentioning that the  in-plane component of the canted magnetic configuration can still retain the octupole order of the triangular AFM state \cite{mn3sn_kerr,mn3sn_octupole_xray}. The coexistence of this octupole order, along with the noncoplanar magnetic order  stabilized by higher-order exchange can lead to a dual order parameter in the system. The transport phenomena of such a magnetic ground state can be  interesting. Motivated by our theoretical results, in this letter,  we experimentally demonstrate the existence of peculiar magnetic and AHE in  the electron-doped Mn$_{3.09}$Sn$_{0.91}$ and Mn$_{3-x}$Fe$_{1+x}$Sn samples. 

%%%%%%%%%%%%%%%%%%%%%%%%%%%%%%%%%%%%%%%%%%%%%%%%%%%%%%%%%%%%%%%%%%%%%%%%%%%%%%%%%%%%%%%%%%%%%%%%%%%%%%%%%%%%

%\section*{RESULTS AND DISCUSSION}
The details of single crystal synthesis and structural characterization are presented in the supplementary material \cite{Supplemantary}. The temperature dependence of magnetization [$ M(T) $] measured with field parallel and perpendicular to the c-axis for  Mn$_{3.09}$Sn$_{0.91}$ and Mn$_{2.7}$Fe$_{0.3}$Sn, and Mn$_{2.5}$Fe$_{0.5}$Sn are depicted in Fig. \ref{Fig1}(d). A distinguishable sharp upturn in the $ M(T) $  data  appears at low temperatures for all the samples when the field is applied parallel to the c-axis. On the contrary, the $ M(T) $ curves for H$\perp$$c$ show very low magnetization due to the near cancellation of the net magnetic moment in the $ab$-plane. The sudden increase in the magnetization at low temperatures for the H$\parallel$$c$  case indicates a realignment of the Mn moments along the c-axis. Most importantly, the temperature corresponding to this spin-reorientation transition ($T_{SR}$) and the out-of-plane magnetic moment  increases  gradually with Fe doping. It is worth mentioning that the $T_{SR}$ for the undoped Mn$_ 3 $Sn sample falls below 50~K  \cite{mn3sn_THE_pradeep}, whereas Mn$_{2.5}$Fe$_{0.5}$Sn exhibits $T_{SR}$=170~K that further increases to near room temperature as observed for the polycrystalline  Mn$_{2.3}$Fe$_{0.7}$Sn sample \cite{Supplemantary}. Further, the large irreversibility between the zero field cooled (ZFC) and the field cooled (FC) $ M(T) $ curves for H$\parallel$$c$   indicates the presence of large out-of-plane magnetic anisotropy. The above results suggest the presence of an insurgent  out-of-plane magnetic moment along with the in-plane triangular AFM state.

To confirm the proposed non-coplanar magnetic ground state in the present system, we next carry out  temperature dependent neutron diffraction (ND) study for the Mn$_{2.5}$Fe$_{0.5}$Sn sample \cite{Supplemantary}. The  350~K ND data recorded above the magnetic ordering temperature can be well fitted by considering the hexagonal Mn$_3$Sn structure \cite{Supplemantary}. The magnetic structure at room temperature is found to be  co-planar inverse triangular in nature \cite{Supplemantary}. To analyze the low-temperature ND data, first we simulate the ND intensity considering the non-coplanar spin ordering, where an appreciable redistribution in the intensity of (200) and (201) reflections is found in comparison to the nearby (102) peak \cite{Supplemantary}. The 3~K ND data with Rietveld refinement using a non-coplanar model is shown in Fig. \ref{Fig1} (e).
%Interestingly, the low temperature ND data exhibit an exact trend to the simulated non-coplanar model, as shown 
The evolution of the non-coplanar structure is also illustrated in Fig. \ref{Fig1} (f).  It can be clearly seen that the intensity of both (200) and (201) peaks starts increasing sharply at the onset of the T$_{SR}$. The enhanced intensities can only be accommodated by considering a non-coplanar canted model of the inverse triangular spin structure. For clarity, we have also plotted the zoomed view of the ND data along with fittings corresponding to the in-plane coplanar and the canted non-coplanar spin structures  around (200) and (201) peaks at selected temperatures [Fig. \ref{Fig1} (g)]. It can be seen that the observed intensities can only be fitted using a canted model below 150~K. The estimated canting angle at 3~K is about 16$^{\circ}$, which decreases to zero at 150~K [inset of Fig. 1(e)].  The full range ND patterns at different temperatures with Rietveld analysis and other fitting parameters are shown in the supplementary material \cite{Supplemantary}. These results categorically establish that the non-coplanar magnetic state is the ground state in the electron-doped Mn$_3 $Sn, hence, go hand in hand with our theoretical model.

%%%%%%%%%%%%%%%%%%%%%%%%%%%%%%%%%%%%%%%%%%%%%%%%%%%%%%%%%%%%%%%%%%%%%%%%%%%%%%%%%%%%%%%%%%%%%%%%%%%%%%%%%%%%%%%%%%%%%%%%%%%%%%%%%%%%%%%%%%%%%
\begin{figure}[tb]
	\begin{center}
		\includegraphics[angle=0,width=8.5 cm,clip=true]{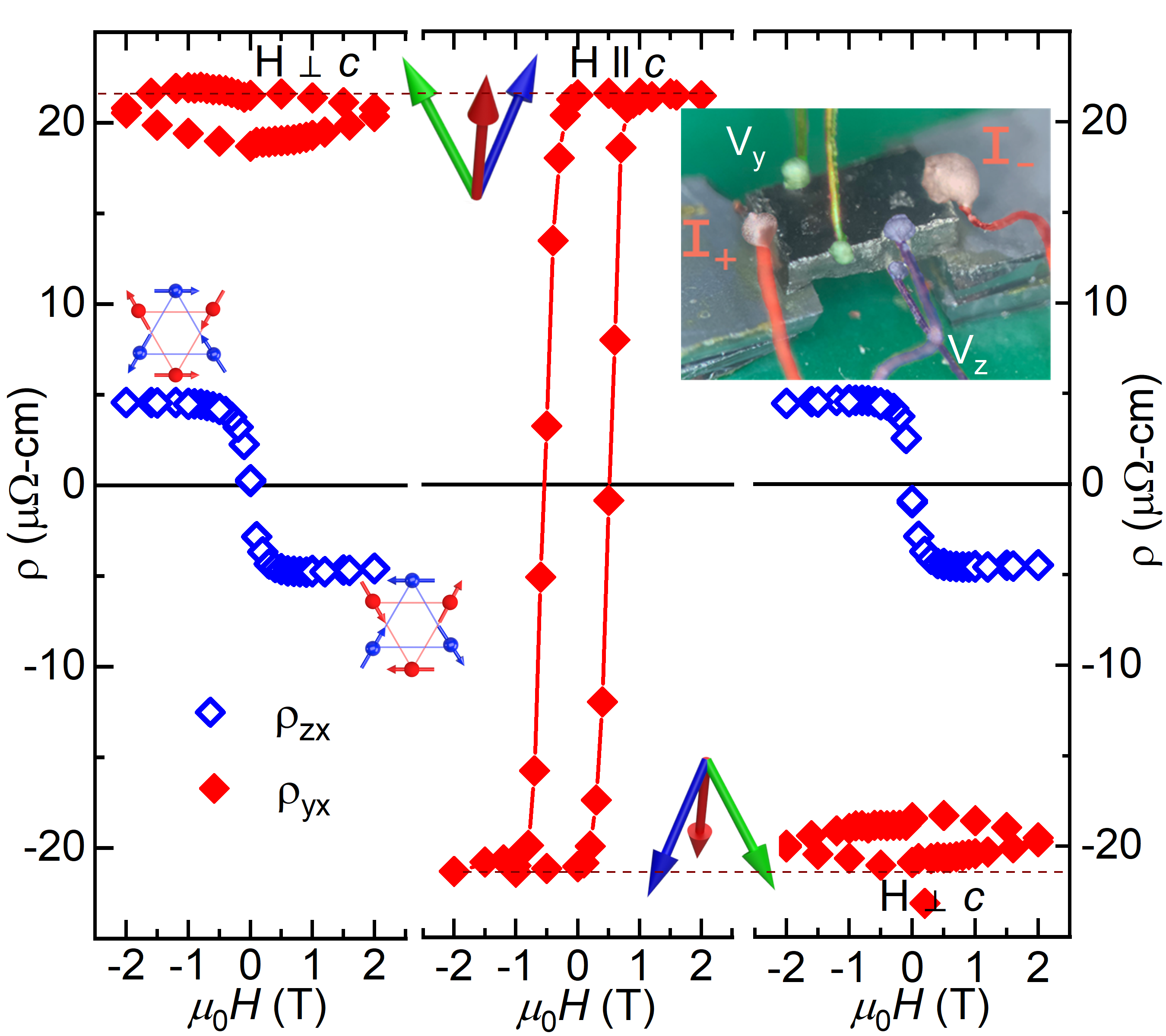}
		\caption{\label{Fig4}  Individual switching of octupole and noncoplanar dipole orders for Mn$_{2.7}$Fe$_{0.3}$Sn measured at 5~K. The middle panel represents  $\rho_{yx}$ with $H \parallel c$. Left and right panels show both $\rho_{yx}$ and $\rho_{zx}$ measurements with $H \perp c$. The image of the sample for the transport measurement  is shown in the inset of the right panel.}
	\end{center}
\end{figure}
%%%%%%%%%%%%%%%%%%%%%%%%%%%%%%%%%%%%%%%%%%%%%%%%%%%%%%%%%%%%%%%%%%%%%%%%%%%%%%%%%%%%%%%%%%%%%%%%%%%%%%%%%%%%

To probe the magnetic and transport properties of the non-coplanar spin state of the electron-doped  Mn$_3 $Sn samples, isothermal magnetization [$ M(H) $] measurements are carried out at 5~K, as shown in Fig. \ref{Fig2}(a-c). All three samples exhibit a large hysteretic behavior  for the $ M(H) $ loops measured with  H$\parallel$$c$, whereas  linear kind of $ M(H) $ curves are found for  H$\perp$$c$. In addition, the saturation magnetization for H$\parallel$$c$  keeps on increasing and reaches  a value of $\approx$2.0$\mu_B$ for Mn$_{2.5}$Fe$_{0.5}$Sn. In the case of the polycrystalline samples, the saturation magnetization further increases to $\approx$4.0$\mu_B$ for Mn$_{2.0}$Fe$_{1.0}$Sn \cite{Supplemantary}. The systematic increase in the out-of-plane magnetic moment further corroborates the evolution of a canted magnetic state from the inverse triangular spin structure with electron doping. 

Next, we perform Hall resistivity measurements on our single crystalline  samples in different geometries to uncover the effect of noncoplanar canted magnetic state on electron transport. The $ M(H) $ loops measured with H$\perp$$c$ clearly show the protection of the 120$^{\circ}$ triangular AFM state,  hence can continue to generate octupole moment, irrespective of the canting that can give nonzero SSC.  It is well known that the octupole  domains of the in-plane AFM configuration align when a field is applied in the  $ab$-plane \cite{mn3sn_kerr}. The  Hall signature corresponding to this order can be observed in the $\rho_{zx}$ component of the Hall resistivity.  The SSC-induced Hall signal can be found from the $\rho_{yx}$ measurement.  Figure \ref{Fig2}(d-f) show the $\rho_{zx}$ and $\rho_{yx}$ data measured at 5~K. All the samples exhibit a robust $\rho_{zx}$ signal that persists up to the magnetic ordering temperature \cite{Supplemantary}, indicating the preservation of the octupole order. Most importantly, a large $\rho_{yx}$ of about 14 $ \mu\Omega $-cm is found for Mn$_{3.09}$Sn$_{0.91}$. With Fe doping,  the $\rho_{yx}$  increases to 25 $ \mu\Omega $-cm for  Mn$_{2.7}$Fe$_{0.3}$Sn and  further surges to an extremely large $\rho_{yx}$ of $ \approx $ 35 $ \mu\Omega $-cm for Mn$_{2.5}$Fe$_{0.5}$Sn. We expect that the $\rho_{yx}$ may attain its maxima for Mn$_{2.25}$Fe$_{0.75}$Sn \cite{Supplemantary}. 

%%%%%%%%%%%%%%%%%%%%%%%%%%%%%%%%%%%%%%%%%%%%%%%%%%%%%%%%%%%%%%%%%%%%%%%%%%%%%%%%%%%%%%%%%%%%%%%%%%%%%%%%%%%%

The $\rho_{yx}$ component of the Hall resistivity  can originate from the out-of-plane component of the magnetic moment that breaks the TRS and the non-zero SSC induced by the non-coplanar magnetic ordering. The $\rho_{yx}$ arising from the trivial out-of-plane magnetic moment must linearly vary with the magnetization.  Whereas the SSC-induced Hall signal depends on the canting angle of the Mn moments and exhibits sinusoidal nature, as shown in Fig. 1(b). Hence, $\rho_{yx}$ originating from   the SSC must depend on the canting angle. Since the canting angle can be changed with the application of a magnetic field, we carry out pulsed field magnetization and  $\rho_{yx}$ measurements up to 60~T on the Mn$_{2.5}$Fe$_{0.5}$Sn sample, as shown in Fig. \ref{Fig3} (a) and (b). The magnetization of the sample increases monotonically with fields up to 60~T for all the measured temperatures [Fig. \ref{Fig3} (a)],  signifying a realignment of the Mn moments along the c-axis. This results in the change of canting angle from about 16$^{\circ}$ at zero field to $\approx$27$^\circ$ at 60~T for $ T= $ 4.2~K. 

To map out the SSC contribution to the $\rho_{yx}$ due to the change in the canting angle, we carry out the 60~T Hall resistivity ($\rho_{yx}$) measurements as shown in Fig. \ref{Fig3} (b). The $\rho_{yx}$  increases with the field and reaches a giant value of $\approx$50 $\mu\Omega $-cm at 60~T for temperature between  4.2~K and 50~K.  Most importantly, $\rho_{yx}$  data exhibit a slower increase at higher fields compared to the magnetization data. This trend can be understood by invoking the SSC contribution to the Hall signal along with the trivial magnetization contribution. According to the scaling relation, the Hall resistivity for the intrinsic contribution (as the present system falls in the bad metal regime) varies as : $\rho_{AHE} \propto \rho^2_{xx} \times M$ \cite{AHE_nagaosa_review,properscaling,AHE_co3sn2s2}. Figure \ref{Fig3} (c) shows the fitting of $\rho_{yx}$ with $\rho^2_{xx} M$. It can be clearly seen that a good fitting is obtained at the low field range, whereas the fitting considerably deviates from the experimental data at very large fields for temperatures between 4.2~K and 100~K. This deviation  at high fields (higher canting angle) originates due to the SSC contribution that becomes more nonlinear at higher canting angles. As expected, a nice fitting with the experimental $\rho_{yx}$ data in the whole field range is obtained at 150~K as the canting angle is almost zero above $T_{SR}$. To makes things more clear, we plot $\rho_{yx}$ as a function of SSC$\times M$ in Fig. \ref{Fig3} (d). In this case, a nice linear fitting is obtained for all temperatures up to 100~K, whereas the fitting deviates for 150~K.  Hence, our high field measurements categorically establish that SSC is the major contributor to the  $\rho_{yx}$ component of the Hall resistivity. Further details of the present analysis can be found in the supplementary materials \cite{Supplemantary}. 

Although we have demonstrated that both $\rho_{zx}$ and $\rho_{yx}$ coexist in the electron-doped Mn$ _3 $Sn samples, it would be really interesting to study if the dual parameters in our samples can be manipulated simultaneously. For this purpose, we have performed AHE measurements on the Mn$_{2.7}$Fe$_{0.3}$Sn single crystal by rotating the sample in a magnetic field. The sample set is prepared in such a way that both $\rho_{zx}$ and $\rho_{yx}$ can be measured at the same time (inset of Fig. \ref{Fig4} right panel). First, the $\rho_{yx}$ component of AHE is measured with $H \parallel c$, where a large SSC-induced AHE signal is clearly visible (Fig. \ref{Fig4}, middle panel). Subsequently, we set the $\rho_{yx}$ to a positive maximum by applying a positive magnetic field before reducing the field to zero.   In this configuration, we rotate the sample to apply in-plane magnetic fields ($H \perp c$)  to measure the octupole order induced AHE, $\rho_{zx}$. At the same time, we also measure the spin chirality induced AHE, $\rho_{yx}$. As it can be seen from the left panel of Fig. 4, the $\rho_{zx}$ can be readily changed from positive to negative value by sweeping the in-plane magnetic fields while keeping the $\rho_{yx}$ at the maximum value. The change in sign of the $\rho_{zx}$ directly confirms a switching of the octupole order, whereas an unchanged $\rho_{yx}$ signifies the preservation of the noncoplanar order. A similar kind of variation in the $\rho_{zx}$  signal is observed by setting the $\rho_{yx}$ at the negative maximum  (Fig. \ref{Fig4}, right panel). We would like to stress here that such individual manipulation of dual parameters is unprecedented in a magnetic system.
 
 % Although the anisotropic AHE has been observed in many materials, the source of this anisotropy is usually the different symmetries of the saturated state when field is applied along different directions. Contrary to our results, the AHE for those cases is expected to go to zero as soon as the magnetic order is aligned along a different direction.

%%%%%%%%%%%%%%%%%%%%%%%%%%%%%%%%%%%%%%%%%%%%%%%%%%%%%%%%%%%%%%%%%%%%%%%%%%%%%%%%%%%%%%%%%%%%%%%%%%%%%%%%%%%%%%%%%%%%%%%%%%%%%

After the discovery of large AHE in the triangular AFM state of Mn$_3 $Sn, a considerable amount of work has been carried out related to the role of underlying magnetic structure in determining the band topology in the system \cite{mn3sn_excitation_park,Weyl_mn3Sn_theory,mn3snWeyl2}. In all the cases, the magnetic properties of the system can be understood by invoking the 2-spin exchange interaction. The present work brings a new direction in the understanding of the role of higher-order interaction  to stabilize the non-coplanar magnetic ground state.  For the present layered kagome lattice structure, the 4-spin and 6-spin terms stabilize a tunable noncoplanar spin structure with dual order characteristics. This  magnetic state gives rise to  a large  Hall signal related to the real space Berry curvature of the system. At the same time, the momentum space Berry curvature induced AHE can be independently manipulated without affecting the former one.  This exemplifies the rich Berry phase physics of the underlying electron band structure in the system. In this direction, it would be really interesting to study the effect of multi-spin interaction to the band topology of the present class of materials.

%%%%%%%%%%%%%%%%%%%%%%%%%%%%%%%%%%%%%%%%%%%%%%%%%%%%%%%%%%%%%%%%%%%%%%%%%%%%%%%%%%%%%%%%%%%%%%%%%%%%%%%%%%%%
%\section*{\textbf{CONCLUSION}}
In conclusion, the present study exemplifies the effects of higher order exchanges in the stabilization of ground state properties of the layered kagome lattice based electron-doped Mn$_3 $Sn. A canted magnetic state  is expected to play the primary role in the realization of a large anomalous Hall signal of topological origin.  In addition to the noncoplanar magnetic state, the octupole order of the layered kagome structure can be preserved for all the samples. This results in a  dual magnetic order, where both these orders can be manipulated individually using magnetic fields.  We hope that the present results will motivate further studies on the effect of higher-order interactions on a variety of lattice classes. In addition, prospective device application of the dual order system can be anticipated.

%%%%%%%%%%%%%%%%%%%%%%%%%%%%%%%%%%%%%%%%%%%%%%%%%%%%%%%%%%%%%%%%%%%%%%%%%%%%%%%%%%%%%%%%%%%%%%%%%%%%%%%%%%%%%%%%%%%%%%%%%%%%%%%%%

\section*{Acknowledgments}
A.K.Nayak acknowledges the Department of Atomic Energy (DAE), the Department of Science and Technology (DST)-Ramanujan research grant (No. SB/S2/RJN-081/2016) and SERB research grant (ECR/2017/000854) of the Government of India for financial support. A.K.Nandy acknowledges the support of DAE and SERB research grant (Grant No. SRG/2019/000867) of the Government of India.  The theoretical computations have been performed on KALINGA  high-performance computing facility at NISER, Bhubaneswar. A.K. Nayak thank Amitabh Das, Bhabha Atomic Research Centre, Mumbai for the recording the initial neutron diffraction data and its preliminary analysis.


\begin{thebibliography}{100}
		
	\bibitem{HOE_Adler_1979} J. Adler and J. Oitmaa, J. Phys. C \textbf{12}, 575 (1979).
	
	\bibitem{3Q_exp_PRL} P. Kurz, G. Bihlmayer, K. Hirai, and S. Bl\"{u}gel, Phys.	Rev. Lett. \textbf{86}, 1106 (2001).
	
	\bibitem{uudd_exp_3spin} A. Kr\"{o}nlein, M. Schmitt, M. Hoffmann, J. Kemmer, N. Seubert, M. Vogt, J. K\"{u}spert, M. B\"{u}hme, B. Alonazi, J. K\"{u}gel, H. A. Albrithen, M. Bode, G. Bihlmayer, and S. Bl\"{u}gel, Phys. Rev. Lett. \textbf{120}, 207202 (2018).
	
	\bibitem{systematicDerivation_highOrder} M. Hoffmann and S. Bl\"{u}gel, Phys. Rev. B \textbf{101}, 024418 (2020).
	
	\bibitem{AHE_spinglass_theory} H. Kawamura, Phys. Rev. Lett. \textbf{90}, 047202 (2003).
	
	\bibitem{AuFeHallSSC} T. Taniguchi, K. Yamanaka, H. Sumioka, T. Yamazaki, Y. Tabata, and S. Kawarazaki, Phys. Rev. Lett. \textbf{93}, 246605 (2004).
	
	\bibitem{AuFeHallSSC2} P. Pureur, F. W. Fabris, J. Schaf, and I. A. Campbell, Europhys. Lett. \textbf{67}, 123 (2004).
	
	\bibitem{AHE_nagaosa_review} N. Nagaosa, J. Sinova, S. Onoda, A. H. MacDonald, and N. P. Ong, Rev. Mod. Phys. \textbf{82}, 1539 (2010).
	
	\bibitem{AHE_SSC_toura_science} Y. Taguchi, Y. Oohara, H. Yoshizawa, N. Nagaosa, and Y. Tokura, Science \textbf{291}, 2573 (2001).
	
	\bibitem{AHE_SSC_withoutdipole} Y. Machida, S. Nakatsuji, S. Onoda, T. Tayama, and T. Sakakibara, Nature (London) \textbf{463}, 210 (2010).
	
	\bibitem{AHE_spinglass_aufe} F. W. Fabris, P. Pureur, J. Schaf, V. N. Vieira, and I. A. Campbell, Phys. Rev. B \textbf{74}, 214201 (2006).
	
	\bibitem{mn3sn_octupole_theory} T. Nomoto and R. Arita, Phys. Rev. Research \textbf{2}, 012045 (2020).
	
	\bibitem{mn3sn_kerr} T. Higo, H. Man, D. B. Gopman, L. Wu, T. Koretsune, O. M. J. van ’t Erve, Y. P. Kabanov, D. Rees, Y. Li, M.- T. Suzuki, S. Patankar, M. Ikhlas, C. L. Chien, R. Arita, R. D. Shull, J. Orenstein, and S. Nakatsuji, , Nat. Photonics \textbf{12}, 73 (2018).
	
	\bibitem{mn3sn_octupole_xray} M. Kimata, N. Sasabe, K. Kurita, Y. Yamasaki, C. Tabata, Y. Yokoyama, Y. Kotani, M. Ikhlas, T. Tomita, K. Amemiya, H. Nojiri, S. Nakatsuji, T. Koretsune, H. Nakao, T. Arima, and T. Nakamura, Nat. Commun. \textbf{12}, 5582 (2021).
	
	\bibitem{Weyl_mn3Sn_theory} H. Yang, Y. Sun, Y. Zhang, W.-J. Shi, S. S. P. Parkin, and B. Yan, New J. Phys. \textbf{19}, 015008 (2017).
	
	\bibitem{AHE_mn3ge_Exp} A. K. Nayak, J. E. Fischer, Y. Sun, B. Yan, J. Karel, A. C. Komarek, C. Shekhar, N. Kumar, W. Schnelle, J. K\"{u}bler, C. Felser, and S. S. P. Parkin, Sci. Adv. \textbf{2}, e1501870 (2016).
	
	\bibitem{AHE_mn3sn_Exp_Nakatsuji} S. Nakatsuji, N. Kiyohara, and T. Higo, Nature (London) \textbf{527}, 212 (2015).
	
	\bibitem{mn3sn_phaseDependHall} N. H. Sung, F. Ronning, J. D. Thompson, and E. D. Bauer, Appl. Phys. Lett. \textbf{112}, 132406 (2018).
	
	\bibitem{mn3sn_pressure} C. Singh, V. Singh, G. Pradhan, V. Srihari, H. K. Poswal, R. Nath, A. K. Nandy, and A. K. Nayak, Phys. Rev. Research \textbf{2}, 043366 (2020).
	
	\bibitem{mn3sn_kondo} D. Khadka, T. R. Thapaliya, S. H. Parra, X. Han, J.Wen, R. F. Need, P. Khanal, W.Wang, J. Zang, J. M. Kikkawa, L. Wu, and S. X. Huang, Sci. Adv. \textbf{6}, eabc1977 (2020).
	
	\bibitem{mn3sn_samplequalityHall} M. Ikhlas, T. Tomita, and S. Nakatsuji, JPS Conf. Proc. \textbf{30}, 011177 (2020).
	
	\bibitem{mn3sn_excitation_park} P. Park, J. Oh, K. Uhl\'{ı}\v{r}ov\'{a}, J. Jackson, A. De\'{a}k, L. Szunyogh, K. H. Lee, H. Cho, H.-L. Kim, H. C.Walker, D. Adroja, V. Sechovsk\'{y}, and J.-G. Park, npj Quantum Mater. \textbf{3}, 63 (2018).
	
	\bibitem{mn3sn_THE_pradeep} P. K. Rout, P. V. P. Madduri, S. K. Manna, and A. K. Nayak, Phys. Rev. B \textbf{99}, 094430 (2019).
	
	\bibitem{Supplemantary} Supplementary.
	
	\bibitem{vasp1} G. Kresse, and J. Hafner, Phys. Rev. B \textbf{47}, 558 (1993).
	
	\bibitem{vasp2} G. Kresse, and J. Hafner, Phys. Rev. B \textbf{49}, 14251 (1994).
		
	\bibitem{vasp3} G. Kresse, and J. Furthm\"{u}ller, Phys. Rev. B. \textbf{54}, 11169 (1996).
	
	\bibitem{vasp4} G. Kresse, and J. Furthm\"{u}ller, Comput. Mater. Sci. \textbf{6}, 15 (1996).
	
	\bibitem{fe3sn}B. C. Sales, B. Saparov, M. A. McGuire, D. J. Singh, and  D. S. Parker, Sci. Rep. \textbf{4}, 7024 (2014).
	
	\bibitem{chiral-chiralExchange} S. Grytsiuk, J.-P. Hanke, M. Hoffmann, J. Bouaziz, O. Gomonay, G. Bihlmayer, S. Lounis, Y. Mokrousov, and S. Bl\"{u}gel, Nat. Commun. \textbf{11}, 511 (2020).
	
	\bibitem{chiralMultisite_Samir} S. Brinker, M. dos Santos Dias, and S. Lounis, Phys. Rev. Research \textbf{2}, 033240 (2020).
	
	\bibitem{properscaling}C. Zeng, Y. Yao, Q. Niu, and H. H. Weitering, Phys. Rev. Lett. \textbf{96}, 037204 (2006).
	
	\bibitem{AHE_co3sn2s2} Q. Wang, Y. Xu, R. Lou, Z. Liu, M. Li, Y. Huang, D. Shen, H. Weng, S. Wang, and H. Lei, Nat. Commun. \textbf{9}, 3681 (2018).
	
%	\bibitem{AHE_scaling_slopeONE} Y. Liu, H. Tan, Z. Hu, B. Yan, and C. Petrovic, Phys. Rev. B \textbf{103}, 045106 (2021).
	
	\bibitem{mn3snWeyl2} J. K\"{u}bler and C. Felser, Europhys. Lett. \textbf{120}, 47002 (2018).
	
\end{thebibliography}
\end{document}